\documentclass{sf2a-conf2012}
\usepackage{graphicx}
\usepackage{hyperref}
\usepackage[]{natbib}  
\usepackage[cyr]{aeguill}
\usepackage{epstopdf}

\def\BibTeX{{\rm B\kern-.05em{\sc i\kern-.025em b}\kern-.08em
    T\kern-.1667em\lower.7ex\hbox{E}\kern-.125emX}}
\bibpunct{(}{)}{;}{a}{}{,}  

\begin{document}

\TitreGlobal{SF2A 2012}


\title{On the role of Alfv\'en waves as precursors of quasi-static acceleration processes in the Earth auroral zone}

\runningtitle{Alfv\'en waves}

\author{F. Mottez}\address{LUTH, Obs. de Paris, CNRS, Universit\'e Paris Diderot, 91190 Meudon}





\setcounter{page}{237}


\maketitle

\begin{abstract}
In the Earth auroral zone, the electron acceleration by Alfv\'en waves is sometimes a precursor of the non-propagating acceleration structures. In order to investigate how Alfv\'en waves could generate non-propagating electric fields, a series of simulations of counter-propagating waves in a uniform plasma is presented. The waves (initially not configured to accelerate particles) propagate along the ambient magnetic field direction. It is shown that non propagating electric fields are generated at the locus of the Alfv\'en waves crossing. These electric fields have a component orientated along the direction of the ambient magnetic field, and they generate acceleration and a significant perturbation of the plasma density. The non-linear interaction of down and up-going Alfv\'en waves might be a cause of plasma density fluctuations (with gradients along the magnetic field) on a scale comparable to those of the Alfv\'en wavelengths. 
\end{abstract}


\section{Introduction}
An auroral substorm is an abrupt increase in night-side auroral power. 
The largest part of the electron acceleration that triggers the auroras happens 
at a few thousands of  kilometres above the ionosphere. Two main families of acceleration processes have been identified: those associated to quasi-static electric fields called \textit{strong double layers}, and those associated to Alfv\'en wave electric fields (see \citep{mottez_2012b} for a review on the role of the Alfv\'en waves). The acceleration by quasistatic, and basically non-propagating, electric fields such as double layers produces mono-energetic beams of electrons, while those by Alfv\'en waves are associated to broadband energy distributions. Recent observations show that these two families of processes are not independent from each other. 

Evidences of acceleration structures emanating from Alfv\'en waves are given by the direct observation of the parallel electric field (parallel to the mean direction of the magnetic field) \citep{Chust_1998, Chaston_2007b} or by the estimate of the wave Poynting flux \citep{Louarn_1994, Volwerk_1996,Keiling_2000}. 

Several papers suggest that Alfv\'enic processes might act as the precursors of quasi-static non-propagating acceleration structures \citep{Zou_2010,Newell_2010,Hull_2010}. 

This paper summaries a series of numerical simulations that investigate the 
interaction of down-going incident Alfv\'en waves with up-going Alfv\'en waves reflected on the ionosphere. Their ability to create parallel stationary electric fields is questioned, as well as to prepare the auroral plasma before the setting of the acceleration processes.

In ideal MHD, Alfv\'en waves do not carry a parallel electric field. Therefore, they cannot accelerate auroral electrons. It has been shown in previous studies \citep{Hasegawa_1975,Goertz_1984} that Alfv\'en waves with an oblique wave vector induce an electric field with a component parallel to the ambient magnetic field. 
In the present study, we don't suppose that the Alfv\'en waves are already accelerating the plasma. As the origin of their small transverse scales is not trivial, it is more careful, for the generality and the simplicity of the initial condition to neglect the presence of small transverse scale associated to the Alvf\'en waves. 

Because of the simple setting of the simulations, the present study cannot pretend to bring any conclusion on the large scale structure of the auroral zone. For instance, we cannot involve the large scale density and magnetic field dependence with altitude, net potential drops on large scales, etc. 
 This paper  focuses on an explanation of the physical process observed in the simulations that are in the range of "microphysics", when compared to magnetospheric scales. Nevertheless, its potential relevance to auroral physics is the main motivation of this work. 
 
{Let us notice already  that we make a distinction between the $X$ component of a vector (parallel to the ambient -and uniform- magnetic field) and its $x$ component, that is parallel to the local value of the magnetic field. 
(The reason why is explained in section \ref{interpretation}.)}

\section{Numerical method and simulation parameters} \label{simulation_parameters}

As in \citep{Mottez_2000, Mottez_2001_b, Mottez_2004_a, mottez_2011c}, dedicated to the physics of auroral Alfv\'en waves, the numerical simulations are made with a EGC (Electron Guiding Center) electromagnetic  PIC code that takes into account the motion of the electron guiding center, and the full ion motion. {The boundary conditions are periodic.} A complete description of this code is given in \citep{Mottez_1998}. The method of initalisation of the Alfv\'en waves is provided in \citep{mottez_2008_a}, and the method for the wave packets in \citet{Mottez_2012a}

The physical variables are reduced to dimensionless variables. Time and frequencies are normalized by the electron plasma frequency $\omega_{p0}$ that correspond to a reference background electron density $n_0$. Velocities are normalized to the speed of light $c$, and the magnetic
field is given in terms of the dimensionless electron gyrofrequency $\omega_{ce}/\omega_{p0}$. The mass unit is the electron mass $m_e$. Therefore, the units   (starting from the Maxwell Eq. in the MKSA system) are 
$c/\omega_{p0}$ for distances, $\omega_{p0}/c$ for wave vectors,  $e$ for charges,
$e n_0$ for the charge density, $c\omega_{ce}/\omega_{p0}$  for 
the electric field.
In the following parts of this paper, all the numerical values and  figures are expressed in this system of units. 

\section{Simulations}
The left-hand side of Fig.1 shows the magnetic field $B_Z$ of two wave packets that propagate in opposite direction.
 The field $B_Z$ is a good proxy of the wave packets positions. We can see that they cross each other at $x \sim 120$ and their intersection starts at $t=200$. 
 The right-hand side of Fig 1 shows the parallel electric $E_X$ field associated to the same wave packets. The alternating fine horizontal lines are associated to the plasma oscillations that are present in any plasmas (their frequency is $\omega_{p0}$). Apart from the plasma waves, we can see that, as predicted by MHD laws, $E_X$ is (almost) null before the intersection. But during the wave packets crossing, it is strongly enhanced. After the intersection, with a small delay, $E_X$ becomes null again.  
 In order to investigate this phenomenons, the problem has been simplified agin. Instead of two wave packets, the crossing of two sinusoidal waves has been studied. 
 The resulting parallel electric field $E_X$ is shown on Fig. 2. Here again, we can see the plasma waves, but also a time independent structure. Because it is present at the start of the simulation, it is not the consequence of an instability (it would not grow instantaneously). Actually, with the two monochromatic waves, the waves interaction starts at the beginning of the simulation since the two waves are present everywhere with the same intensity.
 
{\begin{figure}[ht!]
 \centering
 \includegraphics[width=0.38\textwidth,clip]{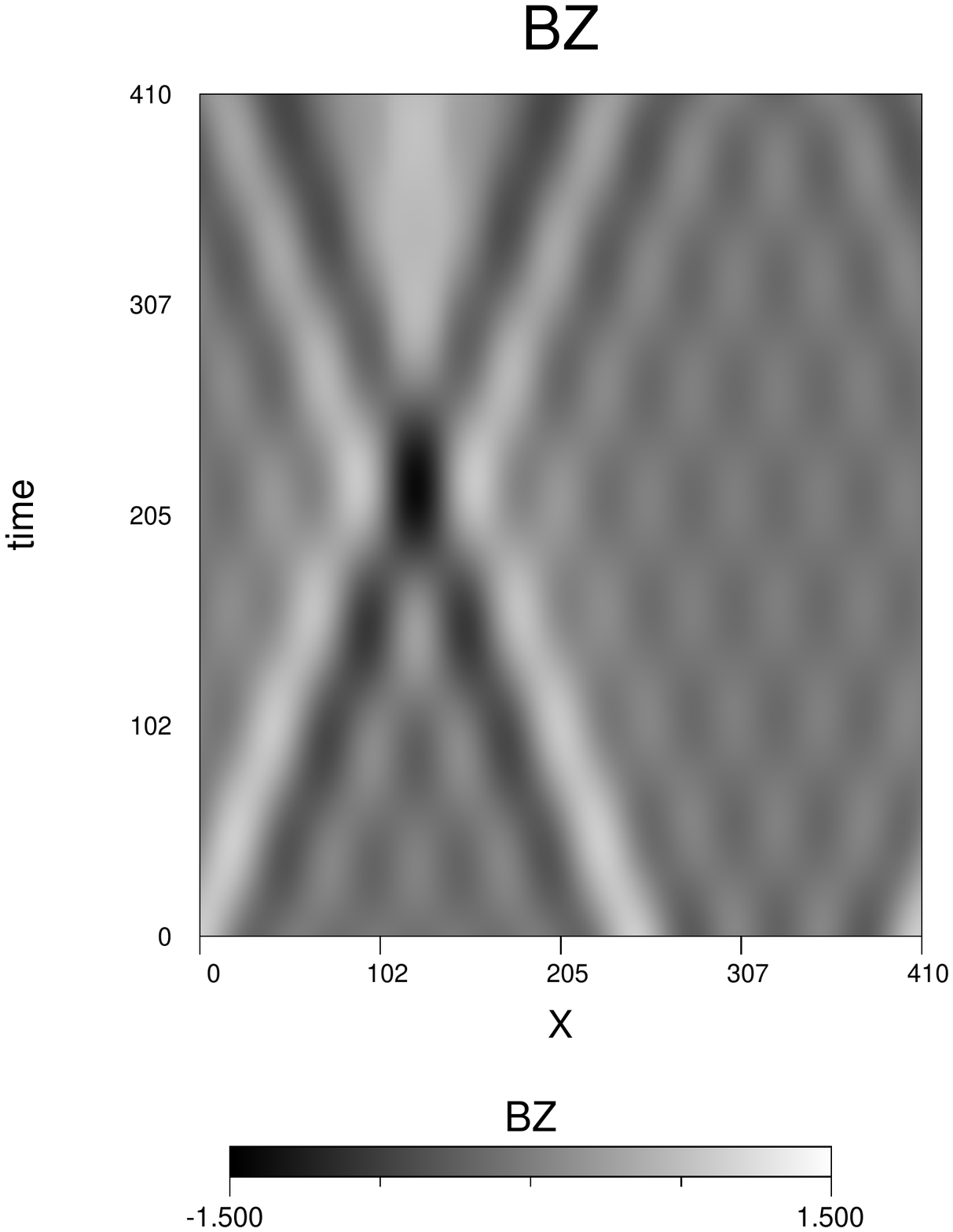}
  \includegraphics[width=0.38\textwidth,clip]{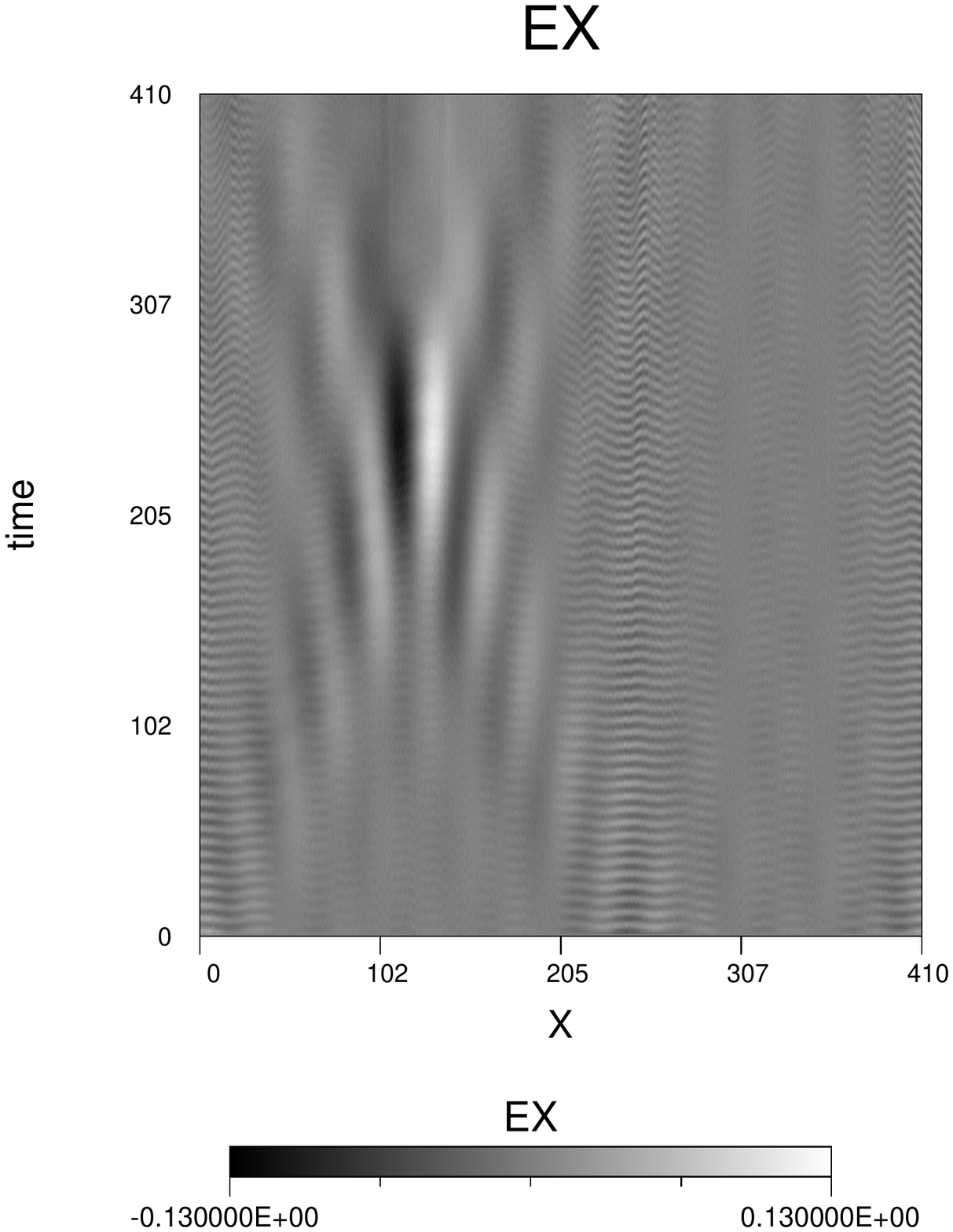}%
  \caption{Plots of electromagnetic field components as a function of position $X$ (horizontal axis) and time $t$ (vertical axis).\textit{Left:} The magnetic field $B_Z$ associated to two wave packets that propagate in opposite direction. \textit{Right:} the parallel electric field $E_X$ associated to the same wave packets.}
\end{figure}

{\begin{figure}[ht!]
 \centering
 \includegraphics[width=0.38\textwidth,clip]{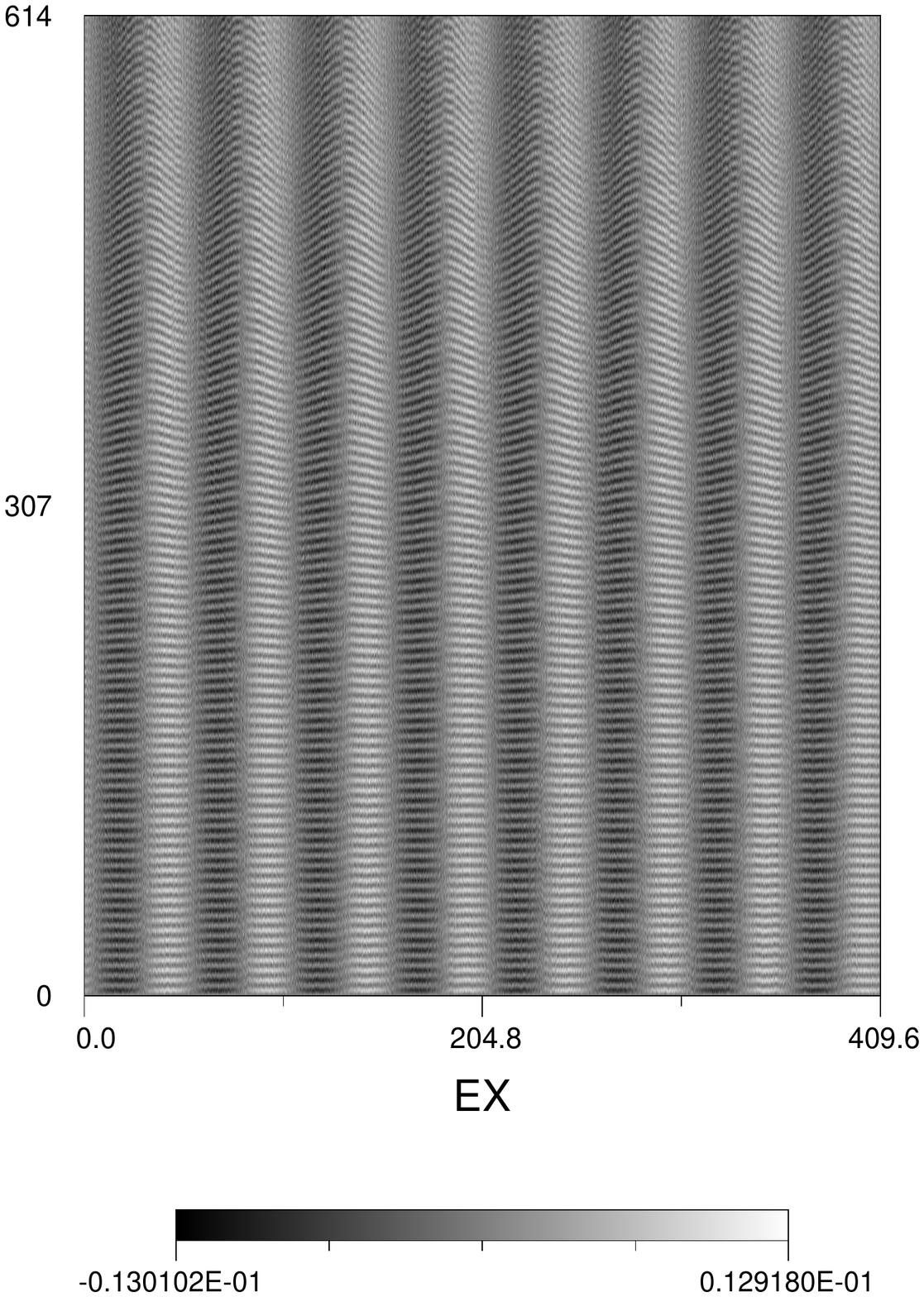}
  \includegraphics[width=0.38\textwidth,clip]{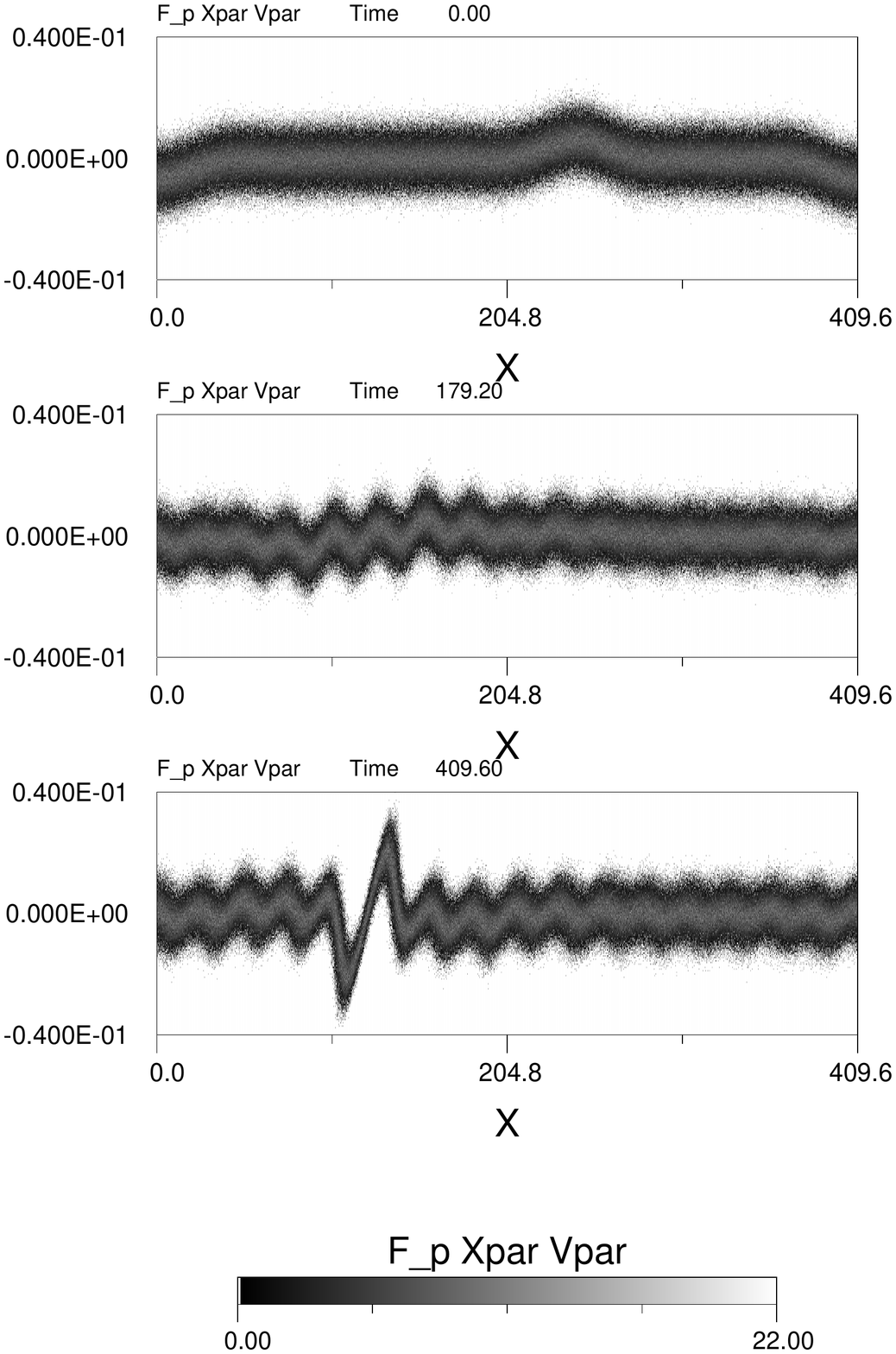}
  \caption{\textit{Left:} The parallel electric field $E_X(X,t)$ associated to a pair of monochromatic Alfv\'en waves that propagate in opposite directions. (Same axis as for Fig. 1). \textit{Right:} The electron phase space density for the same simulation as for Fig. 1. The horizontal axis represents the direction $X$ (it is parallel to the ambient magnetic field), the vertical axis represents the parallel electron velocity $V_X$ and the grey scale is the electron density in that space.}
\end{figure}

\section{How the parallel electric field sets} \label{interpretation}
 An analysis of the properties of $E_X$ was conducted in 
 \citet{Mottez_2012a}, and an explanation was provided. First, it was shown that the intensity of the field is proportional to the product of the intensity of the two sinusoidal waves. Then, it was shown that $E_X$ is always independent of time but dependent on $X$, or independent of $X$ and dependent on time. Then, it was shown that there is no critical value for this phenomenon to appear, it occurs also with low intensity waves, until it is below the noise level. 
 We also projected the electric field along  the local and instantaneous direction $x$ of the magnetic field (that is not uniform because of the perturbation $\delta B_1$ and $\delta B_2$ of the two waves). We found that $E_x=0$.
 Then, it was shown that it could be explained in the following way.
The electric field is assumed to be perpendicular to the direction of the local magnetic field: $\bf E \cdot \bf B=0$. This property is true in the linear theory, it is generalized to non-linear interactions \citep{Knudsen_1996,Tsiklauri_2007}.
Therefore, $E_x=0$, and the projection of the resulting electric field on the $X$ axis can be derived. This derivation (provided in \citet{mottez_2012b}) depends on the polarization of each wave (righ-handed of left-handed circular polarization). It appears that $E_X$ depends of products of $E_1 E_2$, of $E_1^2$ and of $E_2^2$ where $E_1$ and $E_2$ are the amplitudes (set initially) of the two waves.
A second hypothesis is made at this time: we consider that the rule  $\bf E \cdot \bf B=0$ is true as long as it concerns the interaction of two different waves, but not of a wave with itself. (Why ? This point still needs to be clarified.) 
 Then, in the expression of $E_X$, only the terms proportional to $E_1 E_2$ are kept. They perfectly match the properties found in the simulations analysis, according to the various choices of amplitudes, directions of propagation and polarisations.

\section{Particle acceleration and relevance to auroral physics}
The previous analysis shows that the various waves that contribute to a wave packet also interact and they contribute to a parallel electric field (even for a single wave packet). This is why $E_X$ is not strictly null at the begining of the simulation. The figure 3 shows the phase space of the electrons at three different times in the same simulation as for Fig 1. We can see that there is initially a perturbation of the parallel velocity of the electron associated to the initial parallel field $E_X$. 

More interestingly, we can see that after the wave crossing, the parallel velocity is \textit{locally} strongly enhanced, and this enhancement lasts well after the wave packets crossing, contrarily to the field $E_X$. This electron distribution in the phase space presents interesting similarities with those of a newly settle \textit{strong double layer} (localized strong acceleration, well above the thermal level, with a shift of the bulk electron distribution without heating, thus ready to provide quasi mono-energetic electron beams). This is interesting, because as it was said in the introduction, these electrostatic structures (strong double layers) dominate the auroral acceleration after the phase of the Alfv\'enic processes. 

\section{Conclusion}
Two Alfv\'en wave packets crossing each other generate an electric field in a direction $X$ that is parallel to the average ambient magnetic field. This can be explained if we consider that the two waves interact in a way that let the wave electric field and the total magnetic field perpendicular to each other. This is a non-linear wave-wave interaction whose intensity is characterized by the intensity of each wave. The electric field $E_X$ is favourable to electron acceleration, and the phase-space distribution of the electron keeps a signature  of the two waves interaction well after the waves crossing has occurred. The influence of the accelerated electron will be the object of a further study. 

The simulations presented here may provide important clues explaining the transition from the Alfv\'enic to the electrostatic auroral acceleration processes mentioned in the introduction. 



\begin{thebibliography}{20}
\expandafter\ifx\csname natexlab\endcsname\relax\def\natexlab#1{#1}\fi

\bibitem[{{Chaston} {et~al.}(2007){Chaston}, {Hull}, {Bonnell}, {Carlson},
  {Ergun}, {Strangeway}, \& {McFadden}}]{Chaston_2007b}
{Chaston}, C.~C., {Hull}, A.~J., {Bonnell}, J.~W., {et~al.} 2007, Journal of
  Geophysical Research (Space Physics), 112, A05215

\bibitem[{{Chust} {et~al.}(1998){Chust}, {Louarn}, {Volwerk}, {de Feraudy},
  {Roux}, {Wahlund}, \& {Holback}}]{Chust_1998}
{Chust}, T., {Louarn}, P., {Volwerk}, M., {et~al.} 1998, Journal of Geophysical
  Research (Space Physics), 103, 215

\bibitem[{{G{\' e}not} {et~al.}(2000){G{\' e}not}, {Louarn}, \&
  {Mottez}}]{Mottez_2000}
{G{\' e}not}, V., {Louarn}, P., \& {Mottez}, F. 2000, Journal of Geophysical
  Research (Space Physics), 105, 27611

\bibitem[{{G{\' e}not} {et~al.}(2004){G{\' e}not}, {Louarn}, \&
  {Mottez}}]{Mottez_2004_a}
{G{\' e}not}, V., {Louarn}, P., \& {Mottez}, F. 2004, Annales Geophysicae, 6,
  2081

\bibitem[{{G{\'e}not} {et~al.}(2001){G{\'e}not}, {Mottez}, \&
  {Louarn}}]{Mottez_2001_b}
{G{\'e}not}, V., {Mottez}, F., \& {Louarn}, P. 2001, Physics and Chemistry of
  the Earth C, 26, 219

\bibitem[{{Goertz}(1984)}]{Goertz_1984}
{Goertz}, C.~K. 1984, Planetary and Space Science, 32, 1387

\bibitem[{{Hasegawa} \& {Chen}(1975)}]{Hasegawa_1975}
{Hasegawa}, A. \& {Chen}, L. 1975, Physical Review Letters, 35, 370

\bibitem[{{Hull} {et~al.}(2010){Hull}, {Wilber}, {Chaston}, {Bonnell},
  {McFadden}, {Mozer}, {Fillingim}, \& {Goldstein}}]{Hull_2010}
{Hull}, A.~J., {Wilber}, M., {Chaston}, C.~C., {et~al.} 2010, Journal of
  Geophysical Research (Space Physics), 115, 6211

\bibitem[{{Keiling} {et~al.}(2000){Keiling}, {Wygant}, {Cattell}, {Temerin},
  {Mozer}, {Kletzing}, {Scudder}, {Russell}, {Lotko}, \&
  {Streltsov}}]{Keiling_2000}
{Keiling}, A., {Wygant}, J.~R., {Cattell}, C., {et~al.} 2000, Geophysical
  Research Letters, 27, 3169

\bibitem[{{Knudsen}(1996)}]{Knudsen_1996}
{Knudsen}, D.~J. 1996, Journal of Geophysical Research (Space Physics), 10\,761

\bibitem[{{Louarn} {et~al.}(1994){Louarn}, {Wahlund}, {Chust}, {de Feraudy},
  {Roux}, {Holback}, {Dovner}, {Eriksson}, \& {Holmgren}}]{Louarn_1994}
{Louarn}, P., {Wahlund}, J.~E., {Chust}, T., {et~al.} 1994, Geophys. Res.
  Lett., 21, 1847

\bibitem[{{Mottez}(2008)}]{mottez_2008_a}
{Mottez}, F. 2008, Journal of Computational Physics, 227, 3260

\bibitem[{{Mottez}(2012{\natexlab{a}})}]{Mottez_2012a}
{Mottez}, F. 2012{\natexlab{a}}, Annales Geophysicae, 30, 81

\bibitem[{{Mottez}(2012{\natexlab{b}})}]{mottez_2012b}
{Mottez}, F. 2012{\natexlab{b}}, Proceedings of "Waves and Instabilities in
  Space and Astrophysical Plasmas" (WISAP) Eilat, Israel, June 19th - June
  24th, 2011

\bibitem[{{Mottez} {et~al.}(1998){Mottez}, {Adam}, \& {Heron}}]{Mottez_1998}
{Mottez}, F., {Adam}, J.~C., \& {Heron}, A. 1998, Computer Physics
  Communications, 113, 109

\bibitem[{{Mottez} \& {G\'enot}(2011)}]{mottez_2011c}
{Mottez}, F. \& {G\'enot}, V. 2011, Journal of Geophysical Research, 116,
  A00K15

\bibitem[{{Newell} {et~al.}(2010){Newell}, {Lee}, {Liou}, {Ohtani},
  {Sotirelis}, \& {Wing}}]{Newell_2010}
{Newell}, P.~T., {Lee}, A.~R., {Liou}, K., {et~al.} 2010, Journal of
  Geophysical Research (Space Physics), 115, 9226

\bibitem[{{Tsiklauri}(2007)}]{Tsiklauri_2007}
{Tsiklauri}, D. 2007, New Journal of Physics, 9, 262

\bibitem[{{Volwerk} {et~al.}(1996){Volwerk}, {Louarn}, {Chust}, {Roux}, {de
  Feraudy}, \& {Holback}}]{Volwerk_1996}
{Volwerk}, M., {Louarn}, P., {Chust}, T., {et~al.} 1996, Journal of Geophysical
  Research (Space Physics), 101, 13335

\bibitem[{{Zou} {et~al.}(2010){Zou}, {Moldwin}, {Lyons}, {Nishimura},
  {Hirahara}, {Sakanoi}, {Asamura}, {Nicolls}, {Miyashita}, {Mende}, \&
  {Heinselman}}]{Zou_2010}
{Zou}, S., {Moldwin}, M.~B., {Lyons}, L.~R., {et~al.} 2010, Journal of
  Geophysical Research (Space Physics), 115, 12309

\end{thebibliography}

%
\end{document}